\documentclass[prx,twocolumn,amsmath,amssymb,floats,showpacs]{revtex4-2}
\usepackage{graphicx}
\usepackage{bm}
    \usepackage{amsmath}
    \usepackage{amsfonts}
    \usepackage{amsbsy}
    \usepackage{verbatim}
    \usepackage{longtable}
    \usepackage{color}
    \usepackage{tikz}
    \usepackage{braket}
\usepackage[normalem]{ulem}
\usepackage{comment}

\makeatletter
\newcommand{\currentfsize}{\f@size pt}
\makeatother

\renewcommand{\epsilon}{\varepsilon}

\begin{document}

\title{Prethermalization and entanglement dynamics in interacting topological pumps}
%Prethermalization in one-dimensional pumps of interacting particles}

\author{Raffael Gawatz$^1$, Ajit C. Balram$^{1,2,3}$, Erez Berg$^4$, Netanel H. Lindner$^5$, and Mark S. Rudner$^{1,6}$}
\affiliation{$^1$Center for Quantum Devices and Niels Bohr International Academy, University of Copenhagen, 2100 Copenhagen, Denmark \\
$^{2}$Institute of Mathematical Sciences, CIT Campus, Chennai 600113, India \\
$^{3}$Homi Bhabha National Institute, Training School Complex, Anushaktinagar, Mumbai 400094, India \\
$^4$Department of Condensed Matter Physics, Weizmann Institute of Science, Rehovot 7610001, Israel \\
$^5$Physics Department, Technion, 320003 Haifa, Israel\\
$^6$Department of Physics, University of Washington, Seattle, Washington 98195, USA}

\date{\today}

\begin{abstract}
 We investigate the formation of quasisteady states in one-dimensional pumps of interacting fermions at non-integer filling fraction, in the regime where the driving frequency and interaction strength are small compared to the instantaneous single-particle band gap throughout the driving cycle.
 The system rapidly absorbs energy from the driving field, and approaches a quasisteady state that locally resembles a maximal entropy state subject to the constraint of fixed particle number in each of the system's single-particle Floquet bands.
 We explore the nature of this quasisteady state through one-body observables including the pumped current and natural orbital occupations, as well as the (many-body) entanglement spectrum and entropy.
 Potential disorder significantly reduces the amplitude of fluctuations of the quasisteady state current around its universal value, while the lifetime of the quasisteady state remains nearly unaffected for disorder strengths up to the scale of the single-particle band gap.
 Interestingly, the natural orbital occupations and entanglement entropy display patterns signifying the periodic entangling and disentangling of the system's degrees of freedom over each driving cycle.
 Moreover, prominent features in the system's time-dependent entanglement spectrum reveal the emergence of new long timescales associated with the equilibration of many-particle correlations.
\end{abstract}

%\pacs{72.10.Fk, 74.45.+c, 73.63.Kv, 74.50.+r}
\maketitle

\section{INTRODUCTION} \label{Introduction}
Topological phenomena have come to prominence in condensed matter physics due to their robust character, which makes them insensitive to small perturbations and variations of system details~\cite{Stormer1999, Nayak2008, HasanTI_RMP, Qi2011, Yang2015, Ozawa2019}. 
In equilibrium, topological phenomena arise in the ground states (or low-energy excited states) of many-body systems, and therefore require low temperatures for observation.
Interestingly, it has recently been shown that robust topological phenomena may also emerge in far-from-equilibrium quantum many-body systems~\cite{Huse2013,Chandran2014, NandkishoreReview, AFAI, Khemani2016, Else2016, Potter2016, vonKeyserlingk2016, Roy2017, Else2016b, Sacha2017, Parameswaran2018, Balram21d}. 
These non-equilibrium phenomena persist at high energy densities and thus do not require low temperatures for observation.
However, many important questions remain about the factors that control the degree of robustness of these phenomena, as well as their temporal emergence after non-equilibrium driving fields are imposed.

Time periodic driving provides a powerful set of tools for achieving non-equilibrium topological phenomena.
For example, through ``Floquet engineering,'' time-periodic driving may be used to modify the topological properties of Bloch bands in itinerant systems~\cite{Oka2009, Kitagawa2010, Lindner2011, Oka2019, Eckardt2017, RudnerLindnerReview, Cooper2019}. 
However, in the presence of interactions, periodically-driven quantum many-body systems are generically expected to absorb energy from the driving field and heat up towards featureless high entropy density states in which all observables display trivial behavior~\cite{Lazarides2014}.
Disorder-induced many-body localization (MBL) in isolated systems provides a mechanism for inhibiting energy absorption in the system~\cite{Basko2006, Oganesyan2007, Pal2010, NandkishoreReview, Lazarides2015, Ponte2015, Bordia2017}, thereby allowing for a rigorous definition of novel non-equilibrium topological phases that persist in the long-time limit~\cite{Khemani2016, Else2016, Potter2016, vonKeyserlingk2016, Roy2017, Else2016b, Sacha2017, Harper2019}. 
Beyond the strict requirements of MBL and stability in the long-time limit, the use of appropriately designed high-\cite{BukovReview, Eckardt2015, Kuwahara2016, Abanin2017} 
or low-frequency~\cite{Lindner2017} driving fields enables the possibility of studying non-equilibrium phenomena in long-lived transient (``prethermal'') states~\cite{Khemani2020,Ho2020, Fleckenstein2020}.

%%%%%%%%%%%%%%%%%%%%%%%%%%%%%%%%%%%%%%%%%%%%%%%%%%%%%%%%%%%%%%%%%
    \begin{figure}[ht]
	%	\centering
		\includegraphics[scale=1.0]{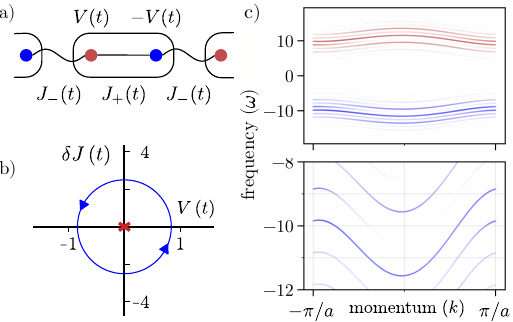} 
		\caption{a) Schematic of a one-dimensional %Rice-Mele 
		quantum pump consisting of a periodically-modulated one-dimensional lattice with a two site unit cell. The sublattice potential $V(t)$ and the intracell and intercell hopping amplitudes,  $J_{+}(t)$ and $J_{-}(t)$, are modulated with a driving period $T$ [see Eqs.~(\ref{eq:model}) and (\ref{eq:drive_para})].  
		b) Driving protocol represented in the plane of dimerization $\delta J(t) \equiv 0.5\left[J_{+}(t) - J_{-}(t)\right]$ and sublattice potential, $V(t)$. The instantaneous band gap closes when $V = 0$ and $\delta J = 0$ (red X).
		The number of times that the drive trajectory encircles the origin of this plane over one cycle (i.e., the winding number) labels topologically-distinct driving protocols.
		c) Time-averaged spectral function of the  
		system in the absence of interactions, for the topologically non-trivial 
		driving protocol indicated by the blue curve in panel b.
		Lower panel: Zoom-in of the lower (right-moving, ``R'') Floquet-band. Each Floquet sideband shifts by $\omega$ as $k$ traverses the Brillouin zone, indicative of a quantized, nonzero value of the band-averaged group velocity. 
		}
		\label{fig:Model}
	\end{figure}
%%%%%%%%%%%%%%%%%%%%%%%%%%%%%%%%%%%%%%%%%%%%%%%%%%%%%%%%%%%%%%%%%
In this work we study the temporal emergence of long-lived 
chiral quasisteady states and associated universal transport behavior in one-dimensional topological pumps, as predicted in Refs.~\cite{Lindner2017, Gulden2019} (see Fig.~\ref{fig:Model}).
We consider a periodically-modulated two-band system, which in the band-insulator regime (i.e., at half-filling) exhibits quantized charge pumping in the adiabatic limit where the modulation frequency is much smaller than the (instantaneous) band gap~\cite{ThoulessPump, Xiao2010, Nakajima2016, Lohse2016}. 
As discussed in Refs.~\cite{Kitagawa2010, RudnerLindnerReview}, 
the corresponding Floquet-Bloch bands of the system are {\it chiral}: quantization of the charge pumped by a filled band arises due to a nonzero average group velocity of the band, which is quantized due to the topological requirement that the quasienergy $\epsilon(k)$ must wind an integer number of times around the quasienergy Brillouin zone as the crystal momentum $k$ traverses the Brillouin zone. 
For a partially-filled band, the pumped charge is non-universal.
However, as discussed in Ref.~\cite{Lindner2017}, in the presence of interactions, the natural heating that is typically viewed as a nuisance for Floquet engineering may actually become a resource: in a low-frequency driving regime where interband transitions are exponentially suppressed in the inverse of the driving frequency, the system is expected to heat up to a state with maximal entropy-density, subject to the constraint of fixed particle number in each band.
In such a state, the occupation of Floquet modes in each band is uniform in $k$, and we expect the quasisteady state current to be given by the topological winding numbers and the corresponding particle densities of the two bands.

While Ref.~\cite{Lindner2017} provided a general picture of this non-equilibrium topological phenomenon, many important open questions remain.
In particular, the extent to which the quasisteady state realizes the heuristic restricted maximal entropy density form described above, as well as how rapidly the current approaches its universal value as a function of system parameters, remain to be addressed.
Moreover, given the central role that disorder plays in the characterization of topological phases, it is important to assess how disorder affects both the fidelity and stability of the quasisteady state of the system.

To assess the degree to which the (constrained) entropy maximization hypothesis is realized, we compute the one-body reduced density matrix, entanglement entropy, and the full entanglement spectrum in the initial transient and quasisteady-state regimes.
Compared with the populations in the system's non-interacting band structure that were studied previously, these observables give a basis-independent view of the quasisteady state that naturally adapts to capture the nature of the steady state even if interaction-induced band renormalizations become large.
Through numerical exact evolution simulations on finite systems, we observe that the  quasisteady state current and natural orbital populations converge to their anticipated forms, while the entanglement spectrum reveals that the full many-body state of the system hosts higher-order correlations that decay only over much longer timescales. 
Remarkably, spatial disorder, which may be expected to assist thermalization by relieving the constraint of momentum conservation (which may particularly severely inhibit scattering in a finite system), significantly improves the convergence of the current to its expected universal value by reducing fluctuations around its mean value. 
However, by all other measures tested, the quasisteady state (including its lifetime) appears to be qualitatively unaffected by the presence of disorder, up to disorder amplitudes comparable to the minimal instantaneous single-particle band gap present within the driving cycle.

In the text below we first describe the model studied, and introduce the Thouless charge pump as described through the framework of Floquet theory. 
We then introduce the observables that we will use 
to study the build-up and nature of the chiral quasisteady state, and discuss the related signatures that we expect. 
Next, we present and discuss our numerical results in relation to the behavior anticipated by our general considerations.
Finally, we discuss open questions and future directions for investigation. 

\section{Problem Setup}
    To study the nature of 
    the quasisteady states that emerge in the low-frequency driving regime, we consider the dynamics of interacting, spinless fermions in a periodically-modulated one-dimensional lattice.
    The full Hamiltonian is given by $H(t) = H_{0}(t)+H_{\rm int} + H_{\rm dis}$, where the translation-invariant single-particle Hamiltonian $H_0(t)$ depends periodically on time, $t$, while $H_{\rm int}$ and $H_{\rm dis}$ describe the interparticle interactions and (static) disorder, respectively. 
    
    The single-particle Hamiltonian $H_{0}(t)$ describes the hopping of fermionic particles between the sites of a one-dimensional lattice with two sublattices, $A$ and $B$ (see Fig.~\ref{fig:Model}a): 
    	\begin{eqnarray}
    		\label{eq:model}
    		%\begin{split}
    		\nonumber H_{0}(t)&=&-J_{+}(t)\sum_{j} c^{\dagger}_{j, A} c_{j, B}-J_{-}(t)\sum_{j} 
    			c^{\dagger}_{j,B}c_{j+1,A} + h.c.\\
    			&&+\ V(t)\sum_{j}\left(c^{\dagger}_{j, A}c_{j, A}-c^{\dagger}_{j, B}
    			c_{j, B}\right),
    		%\end{split}
    	\end{eqnarray}
    where $c^\dagger_{j,s}$ ($c_{j,s}$) creates (annihilates) a fermion in unit cell $j$ on sublattice $s = \{A,B\}$. 
    As depicted in Fig.~\ref{fig:Model}b, the sublattice potential $V(t)$, and intra- and inter-unit cell hopping amplitudes $J_{+}(t)$ and $J_{-}(t)$, respectively, are periodically modulated according to
    	\begin{eqnarray}
    		\label{eq:drive_para}
    		    &V(t)=\delta V_1 \sin(\Omega t),\\
    			\notag &J_{\pm}(t)=J_0 \pm \delta J(t),\quad \delta J(t)=\delta J_0 + \delta J_1 \cos(\Omega t).
    	\end{eqnarray}
    	Here $\delta J_0$ describes a static dimerization of the hopping, and $\delta J_1$ and $\delta V_1$ are the corresponding amplitudes of the modulation at frequency $\Omega$.
    	In the adiabatic limit, at half-filling, this periodically-modulated Rice-Mele system presents a canonical realization of Thouless' quantized adiabatic charge pump~\cite{ThoulessPump, Xiao2010}. 

To study the emergence of quasisteady states, we take a model short-ranged two-body interaction of the form 
\begin{equation}
		H_{\mathrm{int}} = U \sum_j n_j\left(n_j-1\right),\ n_j=c	
		^{\dagger}_{j,A}c_{j,A}+c^{\dagger}_{j,B}c_{j,B}, 
	\end{equation}
	where $U$ is the interaction strength (which we will take to be positive).
	Moreover, we incorporate disorder via a random onsite potential: 
		\begin{equation}
		\label{eq:disorder}
		H_{\rm dis} = \sum_{j,s}\, V_{j,s} \, c^{\dagger}_{j,s} c_{j,s}\,\, , \quad 
		V_{j,s} \in \left[-\eta, \eta\right],
	\end{equation}
	where $V_{j,s}$ is randomly chosen for each unit cell $j$ and sublattice $s = \{A, B\}$ from a uniform distribution over the interval $-\eta \le V_{j,s} \le \eta$, where the parameter $\eta$ encodes the disorder strength.
	We note that quantized pumping for disordered systems in the adiabatic driving regime has previously been shown to be robust to disorder that is weak on the scale of the bulk band gap of the corresponding clean system~\cite{Niu1984, Chern2007, QinJ2016, Wauters2019, Cerjan2020, Nakajima2020, Hayward2020}.

    Due to the time-periodicity of $H_0(t)$, see Eqs.~(\ref{eq:model}) and (\ref{eq:drive_para}), solutions to the single-particle Schr\"{o}dinger equation can be expressed in terms of an orthonormal basis of Floquet states $\ket{\psi_\nu(t)} = e^{-i \epsilon_\nu t}\ket{\phi_\nu(t)}$ that satisfy $i\frac{d}{dt}\Ket{\psi_\nu (t)} = H_0(t)\Ket{\psi_\nu (t)}$. (We set $\hbar = 1$ throughout.) The parameter $\epsilon_\nu$ is the quasienergy of Floquet state $\nu$, and $\ket{\phi_\nu (t)} = \Ket{\phi_\nu (t + T)}$ is periodic in time, with period $T = 2\pi/\Omega$.
	The periodic parts of the Floquet states can be expanded in Fourier ``sideband'' harmonics as
	$\ket{\phi_\nu (t)} = \sum_{m=-\infty}^{\infty}e^{-im\Omega t}\ket{\phi_\nu^{(m)}}$. 
    In the absence of disorder, the single-particle quasienergies form Floquet-Bloch bands, labeled by $\nu = (k,\alpha)$, where $k$ is the crystal-momentum and  $\alpha$ is the Floquet band index.

    The single-particle Floquet band structure of the system we study [Eqs.~(\ref{eq:model}) and (\ref{eq:drive_para})], together with the amplitudes of the sideband harmonics, are conveniently visualized through the time-averaged spectral function~\cite{Usaj2014,FoaTorresMultiTerminal, Uhrig2019, FloquetHandbook}: $\mathcal{A}_{k,\alpha}(\omega) = \sum_{m} A_{k,\alpha}^{(m)} \delta\left(\epsilon_{k, \alpha} + m\Omega - \omega\right)$, with $A_{k,\alpha}^{(m)} = \braket{\phi_{k,\alpha}^{(m)} | \phi_{k,\alpha}^{(m)}}$, see Fig.~\ref{fig:Model}c.
      As displayed in the lower panel, the ``lower Floquet band'' ($\alpha = {\rm R}$ for right-moving, blue color)  carries a quasienergy winding number $w_{\rm R} = 1$: each continuous line of spectral weight shifts up by $\Omega$ as $k$ goes from $-\pi/a$ to $\pi/a$, where $a$ is the lattice constant. The winding number of the upper band ($\alpha = {\rm L}$ for left-moving, red color) is given by $w_{\rm L} = -1$. 
      As a consequence of these nontrivial shifts, the average group velocities of the bands, $\bar{v}_\alpha$, are quantized in units of $a/T$: $\bar{v}_{\alpha} \equiv \frac{a}{2\pi}\int_{-\pi/a}^{\pi/a} dk\, \partial_{k} \epsilon_{k, \alpha} = w_\alpha a/T$.   

In this work, we focus on the situation where the system is initialized with partial filling of the lower (right-mover) band, while the upper (left-mover) band is initially empty. We study the evolution of the interacting many-body system, and characterize its dynamics, quasisteady states, and the emergence of universal charge transport through the observables described below.  

\section{Nature of the quasisteady state} \label{Therm}

	The quasisteady states that we seek emerge due to a parametrically-large separation between the timescales for energy absorption via intraband  and interband scattering, $\tau_{\rm intra}$ and $\tau_{\rm inter}$, respectively.
The populations of particles within the two bands are approximately separately conserved for times $t \ll \tau_{\rm inter}$. 
	For $\tau_{\rm intra} \ll \tau_{\rm inter}$ and on intermediate times $\tau_{\rm intra} \lesssim t \ll \tau_{\rm inter}$, we expect the system to approach a quasisteady state which, 
	from the point of view of all local observables, is equivalent to a state of maximal entropy, subject to the constraint of fixed particle numbers in each band~\cite{Lindner2017}.
	For times $t \gg \tau_{\rm inter}$, we expect the system to approach a featureless state in which all local observables are trivial, as in an infinite temperature state. 

	The separation of timescales $\tau_{\rm inter}/\tau_{\rm intra} \gg 1$ that provides the time window for quasisteady state formation arises when the spectral weights of the two Floquet bands (see Fig.~\ref{fig:Model}c) are well-separated, relative to $\Omega$ and $U$, and to the individual bandwidths. Under these conditions, inter-Floquet-band transitions involve a large energy change that must involve high-order multi-photon absorption or many-body rearrangement processes.

    Before turning to our numerical investigation of this system, we briefly outline the observables that we will use to characterize the evolution and their expected signatures in the quasisteady state.\\

    \noindent \textbf{Period averaged current:} 
   The first diagnostic that we employ is the period-averaged particle current.
   For a system of $N$ unit cells with periodic boundary conditions, the net current flowing around the system, averaged over a driving period starting at time $t_0$, is given by
    \begin{equation}
       \label{eq:current} 
       \mathcal{J}(t_0) = -\int_{t_{0}}^{t_{0}+T}dt\, \dfrac{J_{-}\left(t\right)}{N} \sum_{j=1}^{N} \braket{i c^{\dagger}_{j,B}c_{j+1,A} + h.c.}.
    \end{equation}
    The expectation value in Eq.~(\ref{eq:current}) is taken with respect to the many-body state of the system at time $t$, $\Ket{\Psi(t)}$.
    For a generic initial state, the current at short times takes a non-universal, initialization-dependent value. 
    However, after a short time $\tau_{\rm intra}$, we expect heating via photon-assisted intraband scattering processes to homogenize the populations within the partially-filled band, leading 
    to a universal value of the quasisteady state current: $\mathcal{J}_{\rm qs} = \rho w_{R}/T$, where $\rho$ is the filling fraction of the initially populated $R$ band (recall that we assume the $L$ band to be initially empty), see Fig.~\ref{fig:current_occupation}a.
    On timescales much longer than $\tau_{\rm \rm inter}$, the current will drop to zero as interband scattering causes the band populations equalize. \\ 
    
   \noindent \textbf{Natural orbital populations:}
   More generally, the one-body reduced density matrix (1RDM) $\rho^{(1)}_{ij} = \braket{\Psi(t)|c^{\dagger}_{i} c_{j} |\Psi(t)}$ encodes the expectation values of {\it all} one-body observables.
   Here we focus on the ``natural orbital populations,'' $\{n_{\mu}(t)\}$,  which are the eigenvalues of the 1RDM: $\rho^{(1)} = \sum_{\mu} n_{\mu}(t)\ket{n_{\mu}(t)}\bra{n_{\mu}(t)}$. 
   The natural orbitals $\{\Ket{n_{\mu}(t)}\}$ provide an orthonormal single-particle basis for the system.
   For a translation-invariant state, the natural orbitals are labeled by crystal momentum $k$ and a band index $\tilde\alpha$.
   Compared with the single-particle Floquet bands of $H_0(t)$, the natural orbital bands are renormalized, and 
   incorporate mean-field effects of interactions in the time-dependent state of the system.

    For the reasons above, we track the emergence of the quasisteady state through the natural orbital populations, $\{n_{\mu}(t)\}$,  which provide a more refined picture of the system's dynamics than the non-interacting band populations studied in Ref.~\cite{Lindner2017}.
    For any product (Slater determinant) state, the natural orbital populations are one for each of the filled orbitals, and zero for the remaining unoccupied orbitals. 
    Deviations from this form (of all ones and zeros) signal the development of correlations.
    In the quasisteady state, we expect the natural orbital populations within the (renormalized)
    partially-filled band to be uniform, while the populations in the unoccupied band remain close to zero. 
    For a half-filled band (quarter-filled system), in the quasisteady state we therefore expect $N/2$ natural orbitals (accounting for one of the two bands) to be half-occupied ($n_{\mu} = 0.5$), while the remaining $N/2$ natural orbitals remain nearly empty.
     At long times, $t \gg \tau_{\rm inter}$, we expect all natural orbitals to become equally occupied, with $n_{\mu} = 1/4$ for all $\mu$, as heating allows the system to ergodically explore its full Hilbert space.\\

    \noindent \textbf{Entanglement spectrum and entropy:} 
    While the natural orbital populations provide some measure of correlations between one particle and the remainder of particles in the system, they do not contain sufficient information to resolve correlations of higher order.
    To more fully characterize the many-body nature of the quasisteady state, we turn to the entanglement spectrum and entanglement entropy. 
    These measures characterize entanglement between degrees of freedom that are split across a bi-partition of the system into two complementary regions $\mathcal{A}$ and $\mathcal{B}$. 
    Such correlations are captured by the Schmidt decomposition
        \begin{equation}
            \label{eq:SchmitDecomposition}
    		\ket{\Psi(t)} = \sum_{i} \lambda_{i}\ket{\Phi_{\mathcal{A},i}(t)}\otimes\ket{\Phi_{\mathcal{B},i}(t)};
    \end{equation}
    the entanglement spectrum $\{\xi_i\}$ is defined in terms of the Schmidt coefficients $\{\lambda_i\}$ via $\lambda_i^2 = e^{-\xi_i}$.
    The entanglement entropy is given by the von Neumann entropy of the reduced density matrix of subsystem $\mathcal{A}$ (or $\mathcal{B}$, by symmetry):
    \begin{equation}
    	\label{eq:vNEntropy}
    	S_{\rm vN} = \sum_i \xi_i e^{-\xi_i}.
    \end{equation}
    
    Given that the entropy of a quantum state is preserved under unitary evolution (in particular, remaining zero for any pure state), {\it entanglement entropy} growth due to the build up of correlations plays a crucial role in characterizing self-equilibration and thermalization of closed quantum systems.
    In our context, the tendency of the system to heat towards a state similar to a maximal entropy state, subject to the constraint of fixed band occupations at intermediate times $\tau_{\rm intra} \ll t \ll \tau_{\rm inter}$, is defined precisely in terms of the entanglement entropy and spectrum of a spatially-local subsystem $\mathcal{A}$ of the full system.
    Throughout this work we will consider a ``half-system cut'' bipartition in which subsystems $\mathcal{A}$ and $\mathcal{B}$ are equal in size, each containing $N/2$ unit cells (for a system with a total of $N$ unit cells).

    In the quasisteady state, we expect the state of the system to be similar to a typical random pure state of $N_p$ particles, sampled from the subspace with all particles residing in a single band.
    The corresponding distribution can be described as an ensemble in which all states with $N_p$ particles residing in a single band are represented with equal probability. 
    The corresponding density matrix takes the form of an infinite temperature state (proportional to identity), projected onto the subspace of a single band. 
    Within this ensemble, the von Neumann entropy associated with the reduced density matrix of subsystem $\mathcal{A}$ 
    can be obtained by straightforward combinatorics \cite{Garrison2018}.
    First, the total number of states of the full system, with $N_p$ particles distributed across $N$ unit cells (and restricted to one band), is given by the binomial coefficient $\mathcal{N}_{\rm tot} = {N\choose N_p}$;     the corresponding probability per state is $1/\mathcal{N}_{\rm tot}$.
    For each configuration with $N_p - n$ particles in subsystem $\mathcal{A}$, there are $\mathcal{N}_{\mathcal{B};n} = {N/2\choose n}$  
    configurations of the remaining $n$ particles in subsystem $\mathcal{B}$.
    For a half-filled band, $N_p = N/2$, this yields~\footnote{The superscript $\infty$ in $S_{\rm vN}^{(\infty)}$ denotes that this quantity is calculated for an infinite temperature state, restricted to a single band.}
    \begin{equation}
        \label{eq:S_inf}S^{(\infty)}_{\rm vN} = -\sum_{n=0}^{N/2} 
        \binom{N/2}{N/2-n}\frac{\mathcal{N}_{\mathcal{B};n}}{\mathcal{N}_{\rm tot}} \ln\left(\frac{\mathcal{N}_{\mathcal{B};n}}{\mathcal{N}_{\rm tot}}\right),
    \end{equation}
    where the ratio $\mathcal{N}_{\mathcal{B};n}/\mathcal{N}_{\rm tot}$ is the probability for each state of subsystem $\mathcal{A}$ in which $n$ particles are located in subsystem $\mathcal{B}$, and the factor ${N/2 \choose N/2 - n}$ arises from summing over the (equal) contributions of states with $N/2 - n$ particles in subsystem $\mathcal{A}$.

    Importantly, as pointed out by Page~\cite{Page1993}, the entanglement entropy (and hence entanglement spectrum) obtained by tracing out half of the degrees of freedom of a random pure state, and then averaging over pure states, differs from that obtained for a mixed state corresponding to an ensemble where all states are equally likely (as described above). 
    The ``Page correction'' in the case with a conserved quantity (as we have here for particle number) was further studied in Ref.~\cite{Morampudi2020}.
    For the case of a half-filled band and a half-system cut, in the thermodynamic limit, Ref.~\cite{Morampudi2020} predicts that the entanglement entropy of a random pure state will be reduced by $\Delta S_{\rm vN} = 1/2$ compared with the value $S_{\rm vN}^{(\infty)}$ calculated above for the random ensemble.
    Below we will display both values for comparison when analyzing our numerical results.
   
   Less is known about the  precise form of the entanglement spectrum for a random pure state in the restricted Hilbert space.
    However, based on the similarity to the constrained maximal entropy mixed state discussed above, we may expect that entanglement spectrum in the quasisteady state will feature $\sim N_p/2$ clusters of approximately degenerate entanglement levels, corresponding to the states with $N_p - n$ particles in subsystem $\mathcal{A}$, with degeneracies given by the weight factors $\mathcal{N}_{\mathcal{B};n}$ defined above.

\section{Numerical Simulations} \label{Results}	
	
We now discuss our numerical results for the one-body and many-body observables described in the preceding section. 
    Except when noted otherwise, all numerical results were obtained by exact time-evolution of a periodic system of $N=16$ unit cells ($L = 32$ sites) with $N_{p} = 8$ fermions (quarter filling). 
	In each simulation the system is initialized in 
	a single Slater determinant state. 
	For the clean systems (i.e., with no disorder), 
	we take an initial state in which all single particle states with positive values of crystal momentum in the lower Floquet band are occupied. 
	(This band is adiabatically connected to the lower band of the original, non-driven system~\cite{footnote:initialization}.)
	In the presence of disorder, we initialize the system by filling the $N_{p}$ single particle Floquet eigenstates of the disordered system that have the largest projections on the lower Floquet band of the clean system.
	Although the initial transients depend on the details of the initialization, we observe the behavior in the quasisteady and long-time regimes to be insensitive to the initialization as long as the initial band populations are kept fixed.\\

	%%%%%%%%%%%%%%%%%%%%%%%%%%%%%%%%%%%%
	  \begin{figure}[t]
		\centering
		\includegraphics[width=0.45\textwidth]{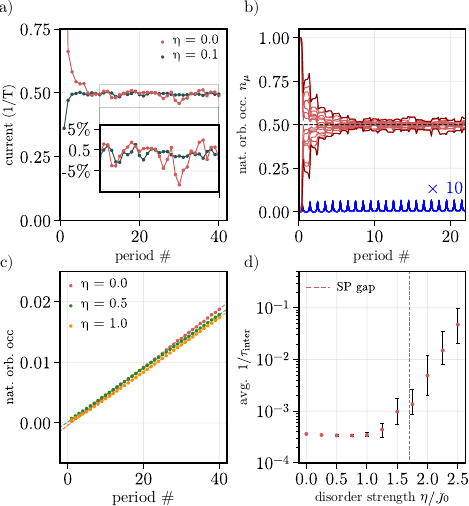}
		\caption{Transient and quasisteady state behavior of single-particle observables. 
		a) Period-averaged current for clean and disordered systems, at quarter filling. The system is initialized with the right-moving Floquet band half-filled. The current starts from an initialization-dependent value,  then quickly approaches the expected universal value [see discussion below Eq.~(\ref{eq:current})]. Inset: zoom-in showing deviation of the period-averaged current from the quasisteady state value. Fluctuations are significantly suppressed by disorder. 
		b) The populations of the 16 natural orbitals with highest occupation (approximately corresponding to the right-moving Floquet band) quickly converge around the expected value of 1/2 (red curves);  the remaining 16 nearly-empty natural orbitals are shown in blue, magnified by a factor of $10$ for visibility. c) Total occupation of the initially empty band, measured by the sum of the 16 smallest natural orbital populations. 
		The timescale for interband excitation defined by the slopes of these curves ($\tau^{-1}_{\mathrm{inter}} \approx 3 \times 10^{-4}/T$) is nearly unaffected by disorder.
		d) Effect of disorder strength $\eta$ on the 
		interband excitation rate $\tau^{-1}_{\mathrm{inter}}$.  
		Each point is an average taken over 50 disorder realizations in a system of $L=28$ sites.
		The red dashed line indicates where the disorder bandwidth $2\eta$ is equal to the minimal instantaneous single particle band gap over the driving cycle.
		Drive parameters (also corresponding to Fig.~\ref{fig:Model}c): $\Omega=0.23,\, J_0=1.0,\, \lambda=0.85,\, \delta J_0=0.0,\, %V_0=0.0,\, 
		U = 2J_0$. 
		The modulations of the hopping and sub-lattice potential are set by $\delta J_1 = \lambda J_0 $ and $V_1=3\lambda J_0$.}
		\label{fig:current_occupation}
	\end{figure}
	%%%%%%%%%%%%%%%%%%%%%%%%%%%%%%%%%
	\noindent{\bf One-body Observables:}
	Time traces of the current and natural orbital occupations in a system in the topologically-nontrivial driving regime are shown in Fig.~\ref{fig:current_occupation} (see caption for parameter values).
	As anticipated, the current quickly approaches and then fluctuates around the universal value of $0.5/T$ characteristic of a half-filled band with winding number $w_{\rm R} = 1$ in the quasisteady regime (panel a).
	On a similar timescale $\tau_{\rm intra}$ of a few driving periods, the populations of 16 natural orbitals coalesce around the value 0.5, while the remaining 16 natural orbital populations remain close to zero, see Fig.~\ref{fig:current_occupation}b. (Here $N = 16$ is the number of unit cells of the system.)
	We have confirmed that the 16 natural orbitals with populations near 0.5 have nearly unit overlap with the states in the right-moving band of the system in the absence of interactions.
	Interestingly, the populations of the remaining 16 natural orbitals display periodic spikes (repeating once per driving cycle).
	These spikes indicate that, within each driving cycle, the system builds up -- and then eliminates -- additional correlations that cannot be described by projecting into {\it any} single band.
	We expect that such reversible interband mixing by inter-particle interactions arises due to the periodic modulation of the instantaneous band gap and interaction matrix elements throughout the driving cycle.
	Interestingly, in this topologically nontrivial regime the times of maximal mixing correspond to times where the instantaneous Wannier orbitals straddle the bonds between adjacent unit cells (cf.~discussion of entanglement entropy below); we surmise that an enhanced interaction probability at these times is thus responsible for the spike.

	In the presence of disorder, which relieves the conservation of crystal momentum, the fluctuations of the current around the quasisteady state value of $0.5/T$ are dramatically reduced (Fig.~\ref{fig:current_occupation}a).
	To explain this behavior, we first note that for a clean system, time-dependent deviations of the current from the anticipated universal value arise due to non-uniform (instantaneous) population of the Floquet states in the right-mover Floquet band. 
	Disorder helps to reduce such fluctuations by mixing and causing more efficient scattering between states from across the Brillouin zone, thus providing more uniform sampling of the subspace that is on average transported by one unit cell to the right per driving period.
	Indeed, along with the reduction of fluctuations of the current, we observe that the distribution of natural orbital populations is more homogeneous and temporally stable in the presence of disorder than in the clean case.
	Given that disorder appears to improve the quality of the quasisteady state from the point of view of its transport properties,
	it is natural to ask how disorder affects other aspects of the quasisteady state, including its stability in particular.

	We now further characterize the role of disorder in the processes that lead to the eventual decay of the quasisteady state.
	In Fig.~\ref{fig:current_occupation}c we plot the total occupation of the initially empty band of natural orbitals, as a function of time.
	After a small initial jump at short times, the excited population grows approximately linearly in time; the slope of this linear dependence gives a measure of the decay rate~\cite{Lindner2017} $\tau_{\rm inter}^{-1}~\sim 10^{-4}/T$.
	In Fig.~\ref{fig:current_occupation}d we plot the (disorder-averaged) interband excitation rate $\tau_{\rm inter}^{-1}$, extracted from such linear fits, as a function of disorder strength.
	Here we see that the decay rate remains nearly unaffected by disorder up to quite large disorder strengths (of order the band width), with a breakdown occuring once the disorder strength becomes comparable to the minimal instantaneous single-particle gap (red dashed line).
	Thus we find that disorder dramatically improves the degree of quantization of the quasisteady state current, without limiting its lifetime.
    
    We observe qualitatively similar behavior in the topologically trivial driving regime, albeit with the current decaying to zero over the time $\tau_{\rm intra}$ (see data in Appendix \ref{app:TrivialRegime}).
    In particular, similar to Fig.~\ref{fig:current_occupation}b, the populations of the nearly empty natural orbitals exhibit periodic spikes.
    Interestingly, in the topologically-trivial driving regime, the spikes show additional structure that closely correlates with modulations of the instantaneous gap between single-particle bands that occur throughout the driving cycle, with maximal interband correlations appearing at times coinciding with minima of the instantaneous band gap.\\

%%%%%%%%%%%%%%%%%%%%%%%%%%%%%%%%%
%% Entanglement figure 
	\begin{figure}[!h]
		\includegraphics[width=0.45\textwidth]{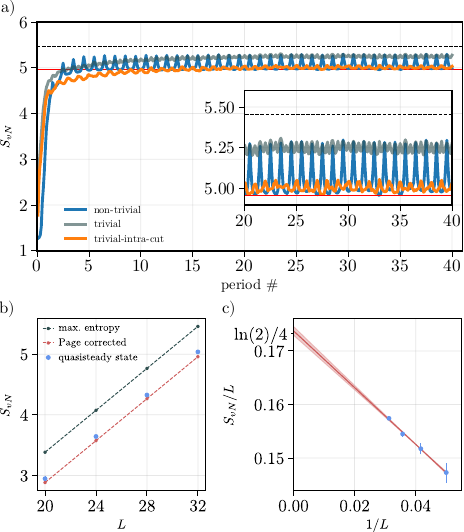} 
		\caption{
    		Time evolution of the entanglement entropy $S_{\rm vN}$ [Eq.~(\ref{eq:vNEntropy})] and 
    		scaling of its saturation value in the quasisteady state with number of sites, $L$. 
        		a) Entanglement entropy $S_{\rm vN}$ for a half-system cut	for a clean system in the topologically trivial (gray and orange curves) and non-trivial (blue curve) driving regimes.
        		The black dashed and red solid lines indicate the entanglement entropies of the band-projected infinite-temperature state, $S^{(\infty)}_{\rm vN}$ [Eq.~(\ref{eq:S_inf})], and its Page-corrected value, respectively.
        		When the entanglement partition cuts through the tails between Wannier centers in the populated band, the quasisteady state entanglement entropy of the topologically trivial system sits just above the Page-corrected value (orange curve).
        		For an entanglement partition that cuts through the maximum of the Wannier function, an additional
        		contribution to $S_{\rm vN}$ is observed in the quasisteady state (gray curve).
        		Periodic spikes observed for the topologically non-trivial drive (blue curve) arise due to polarization pumping across the entanglement cut during the driving cycle.
        		b) Scaling of the absolute difference between the average of the time-averaged  
        		entropy $S_{\rm vN}$ in the numerically obtained quasisteady state (between the 10th-20th period) and $S^{(\infty)}_{\rm vN}$, 
        		with system size, $L$ (at quarter filling). 
        		c) Extrapolation of the entanglement entropy density $S_{\rm vN}/L$ to the thermodynamic limit is consistent with the expected value of $\ln (2)/4$ for an infinite temperature state.
        		Error bars show the standard deviation of the period-averaged entanglement entropy for periods 10-20.
    		Driving parameters for the topologically-trivial regime:  $\Omega=0.1,\, J_0=3.5,\, \lambda=0.25,\, \delta J_0=-2.2J_0,\, %V_0=0.0,\, 
    		U =1.0J_0$. Parameters for the topologically nontrivial driving regime are as in Fig.~\ref{fig:current_occupation}.
		}
		\label{plot:entropy}
	\end{figure}
\begin{figure}[t]
		\centering
		\includegraphics[width=0.45\textwidth]{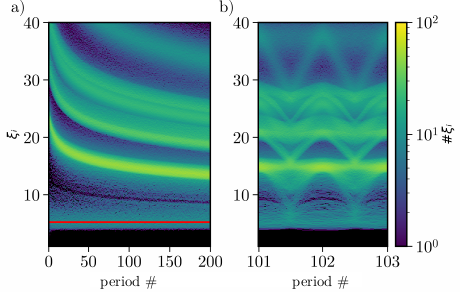}
		\caption{Histogram of the time-dependent entanglement spectrum,
		$\{\xi_i\}$ [see text below Eq.~(\ref{eq:SchmitDecomposition})].  
		a) Evolution of the stroboscopic ES histogram at integer multiples of the driving period.  
		Bright streaks showing accumulation of entanglement levels  slowly converge towards $\xi_{\rm qs}$, signifying the emergence of new long relaxation timescales associated with relaxation of many-body correlations. b) Inter-period evolution of the ES histogram. Sloped lines running between the bright streaks are associated with winding of the ES in the pumping regime \cite{Hayward2018}.
		Driving parameters are the same as in Fig.~\ref{fig:current_occupation}, with $\eta = 0$.}
		\label{plot:es_spectrum}
	\end{figure}

%%%%%%%%%%%%%%%%%%%%%%%%%%%%%%%%%
	\noindent\textbf{Entanglement diagnostics:}
    In Fig.~\ref{plot:entropy} and Fig.~\ref{plot:es_spectrum} we show how the entanglement entropy and entanglement spectrum evolve in time.
    Here we focus on a clean system; results for disordered systems differ only in small quantitative details.
    Similar to the rapid equilibration of the one-body observables displayed in Fig.~\ref{fig:current_occupation}, the entanglement entropy $S_{\rm vN}$ [Eq.~(\ref{eq:vNEntropy})] for a half-system cut rapidly approaches a quasisteady saturation value over the short timescale $\tau_{\rm intra}$ of a few driving periods (Fig.~\ref{plot:entropy}a). 
    For driving parameters in the topologically nontrivial regime (blue curve), $S_{\rm vN}$ displays periodic spikes akin to (and in phase with) those seen for the natural orbital populations.
    In the topologically trivial driving regime, aperiodic fluctuations with a much smaller amplitude are observed (gray and orange curves, see inset for zoom-in).

   The black dashed line in Fig.~\ref{plot:entropy} shows the value $S^{(\infty)}_{\rm vN}$ obtained from an infinite temperature state projected into a single band [Eq.~(\ref{eq:S_inf})]; the red horizontal line below it is offset by the expected Page correction~\cite{Page1993,Morampudi2020}, $\Delta S_{\rm vN} = 0.5$.
    The orange and gray traces were obtained for a system in the topologically-trivial driving regime, with two choices of entanglement cuts (either both cutting through bonds with hopping $J_+$, or $J_-$, see Fig.~\ref{fig:Model}a). 
    For the gray curve, the partition cuts through the Wannier orbital of the occupied band near its maximum, giving an additional contribution to the entanglement entropy for the particles that straddle the cut. 
    For the orange curve, the partition cuts through the tails between Wannier centers, and we see that the saturation value of the entanglement entropy in the quasisteady state falls very close to the Page-corrected value anticipated above.
    The spikes in the blue curve (topologically nontrivial driving regime) interpolate between these two values as the Wannier centers periodically sweep past the entanglement cuts due to the nontrivial pumping in this regime.

   In Figs.~\ref{plot:entropy}b and c, we show the scaling of the saturation value of the entanglement entropy with the number of sites in the system, $L$.
   The data shown are for the topologically nontrivial driving regime, and are averaged over the last 10 periods of each run to smooth out the periodic spikes.
   The small black points show the corresponding values of $S^{(\infty)}_{\rm vN}$ for each system size (as indicated by the black dashed line in panel a for the system with $L = 32$ sites).
   The small red points indicate the corresponding Page-corrected values.
   In Fig.~\ref{plot:entropy}c we plot the entropy density as a function of $1/L$. 
   The data extrapolate well to the expected value of $\ln(2)/4$, corresponding to a half-filled band, in the thermodynamic limit.

    In Fig.~\ref{plot:es_spectrum}a we show a  histogram (color scale) of the half-system-cut entanglement spectrum of the system at integer multiples of the driving period, $T$;
    the color scale can be interpreted as an ``entanglement density of states.''
    A zoom-in showing the entanglement spectrum at intermediate times over two complete periods in the quasisteady regime is shown in Fig.~\ref{plot:es_spectrum}b.
        
    As discussed below Eq.~(\ref{eq:S_inf}), in the quasisteady state we expect the entanglement levels to approximately concentrate around several values, distinguished by the numbers of particles in subsystems $\mathcal{A}$ and $\mathcal{B}$.
    The states with approximately equal numbers of particles in subsystems $\mathcal{A}$ and $\mathcal{B}$ are by far the most probable, and therefore appear with largest Schmidt coefficients [see Eq.~(\ref{eq:SchmitDecomposition})].
    The corresponding entanglement levels appear with the lowest values of $\xi_i$, and provide the dominant contribution to the entanglement entropy, $S_{\rm vN}$ [Eq.~(\ref{eq:vNEntropy})].
    For the reference case of an infinite temperature state projected into one band, the entanglement density of states would consist of a set of bright peaks at the values $\ln (\mathcal{N}_{\mathcal{B};n}/\mathcal{N}_{\rm tot})$ with weights given by the corresponding combinatorial factors in Eq.~(\ref{eq:S_inf}).
    The horizontal red line in Fig.~\ref{plot:es_spectrum}a indicates the value $\ln (\mathcal{N}_{\mathcal{B};N_p/2}/\mathcal{N}_{\rm tot})$, which would correspond to the brightest peak in the infinite temperature state.

    The plots in Figs.~\ref{plot:es_spectrum}a,b display a much richer structure than anticipated for the simple picture described above.
    First, despite the fact that the one-body observables are very close to their expected values, the entanglement spectrum displays several ``streaks'' (of low total weight) around which many entanglement levels are concentrated. (While there is significant weight near the red line, which accounts for the bulk of the entanglement entropy, there is no visible peak structure associated with the different particle number sectors as anticipated above for the band-projected infinite temperature state.)
    The bright streaks of low Schmidt weight slowly shift towards lower values of $\xi$ [and hence towards greater weight in the Schmidt decomposition, Eq.~(\ref{eq:SchmitDecomposition})], indicating that their (small) contributions to the quasisteady state increase with time.
    The time-dependencies of the streak positions are neither described by exponential nor any simple power law forms. 
    This behavior indicates the emergence of additional long timescales, between the timescale $\tau_{\rm intra}$ for equilibration of one-body observables to the quasisteady regime and the timescale $\tau_{\rm inter}$ for relaxation of the band populations.
    These additional long timescales for equilibration within the band can be attributed to a slow build-up of many-body correlations.

	The intra-period evolution of the entanglement spectrum (ES) showcased in Fig.~\ref{plot:es_spectrum}b displays an interesting oscillatory pattern. 
    On a heuristic level, the periodic entangling/disentangling of the system across its partition can be traced to the change of polarization that occurs during a nontrivial pumping cycle~\cite{Hayward2018, Zaletel2014}. 
    Indeed, by labeling each entanglement level by the corresponding number of particles in subsystem $\mathcal{A}$ in the Schmidt decomposition of the state [see Eq.~(\ref{eq:SchmitDecomposition})], the diagonal lines connecting between the bright horizontal bands of entanglement levels in Fig.~\ref{plot:es_spectrum}b can be traced to the movement of particles across the partition. 
    This behavior is responsible for the periodic spikes observed in the $S_{\rm vN}$ in Fig.~\ref{plot:entropy}a.
    Note that these features are not observed in the topologically-trivial driving regime (not shown).

\section{Discussion} \label{Discussion}
In this work, we have provided a multifaceted view of prethermalization and the formation of chiral quasisteady states in many-body quantum pumps. 
From the point of view of one-body observables, we identify a clear intraband thermalization timescale $\tau_{\rm intra}$ that characterizes the approach of these observables to quasisteady state values.
Remarkably, the addition of weak potential disorder {\it reduces} fluctuations of the period-averaged current around its expected universal value.
This highlights not only the robustness of universal far from equilibrium topological transport, but also showcases the positive role that disorder can play in reducing the fluctuations of the current around its universal value by helping to homogenize the state of the system across the topologically-transported subspace. 

The quasisteady state persists for disorder strengths exceeding the single particle bandwidth, while still remaining smaller than the instantaneous single-particle band gap.
The lifetime $\tau_{\rm inter}$ of the quasisteady state, as measured by the growth rate of population in the initially nearly-empty band of natural orbitals, is similarly only very weakly affected by disorder until the disorder strength approaches the single-particle band gap.
Interestingly, in the absence of interactions, a topological transition to a trivial phase (with all particles remaining localized over each driving cycle) is expected when the disorder strength is made larger than single-particle gap~\cite{Niu1984, Wauters2019, Nakajima2020}.
It will be an interesting direction for future work to study the interplay (and possible relation) between this disorder-driven topological transition and the destruction of the quasisteady state due to fast interband heating.

Our study of the entanglement spectrum of the system also revealed intriguing structure as well as the emergence of long equilibration timescales, intermediate between $\tau_{\rm intra}$ and $\tau_{\rm inter}$.
This structure highlights that, even with rapid absorption of energy from the driving field, the formation of a maximal-entropy-like state (restricted to the low energy sector) is a complex process that requires the development of many-particle correlations. 
Even at times when one-body observables such as the current have already converged to their expected quasisteady state values, many-particle correlations continue to develop and evolve over long times.
Further investigation of this structure and slow equilibration of many-body correlations, along with their implications for the reversibility of the system's dynamics, provide additional fertile ground for further investigation.\\

\noindent\textbf{Acknowledgements:} R.~G., A.~C.~B., and M.~R. gratefully acknowledge the support of the European Research Council (ERC) under the European Union Horizon 2020 Research and Innovation Programme (Grant Agreement No.~678862), and the Villum Foundation.  E.~B.~and M.~R. acknowledge support from CRC 183 of the Deutsche Forschungsgemeinschaft (project A01). N.~L.~acknowledges support from the European Research Council (ERC) under the European Union Horizon 2020 Research and Innovation Programme (Grant Agreement No. 639172).
	
\appendix

\section{Additional data for the topological trivial driving regime}
\label{app:TrivialRegime}
\begin{figure}[!ht]
	    \includegraphics[width=0.96\columnwidth]{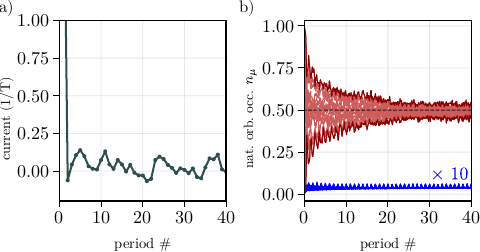} 
	    \caption{One-body observables in the topologically-trivial driving regime. 
	    a) As expected for the trivial phase, the current decays to zero over the intraband equilibration timescale $\tau_{\rm intra}$.
	    b) Similar to the behavior in the topologically nontrivial driving regime shown in the main text, the 16 highest natural orbital populations converge to values around 0.5 on the timescale $\tau_{\rm intra}$. The remaining 16 natural orbital populations show periodic spikes (magnified by a factor of 10), with structure that correlates well with the size of the instantaneous single-particle band gap.
	    Simulation parameters: $\Omega=0.1,\, J_0=3.5,\, \lambda=0.25,\, \delta J_0=-2.2 J_0,\, U =1.0J_0$.}
	    \label{fig:appOneBody}
	\end{figure}

In this appendix, we provide additional data for the topologically-trivial driving regime, to complement the data presented in the main text.
In Fig.~\ref{fig:appOneBody}, we show the time dependence of the current and natural orbital populations.
The behavior is qualitatively similar to that observed for the topologically-nontrivial driving regime, as shown in Fig.~\ref{fig:appOneBody} of the main text, albeit with the current decaying to zero over the time $\tau_{\rm intra}$ as expected for the topologically-trivial regime.
%%%%%%%%%%%%%%%%%%%%%%%%%%%%%%%%%%%%%%%%%%%%%%%%%%%%%%%%%%%		
%%%%%%%%%%%%%%%%%%%%%%%%%%%%%%%%%%%%%%%%%%%%%%%%%%%%%%%%%%%

\bibliographystyle{apsrev4-1}
\bibliography{paper_prethermal}

\end{document}